\documentclass[conference]{IEEEtran}

\IEEEoverridecommandlockouts                              

\usepackage{graphicx}
\usepackage[utf8]{inputenc}
\usepackage{url}

\usepackage{balance}  
\usepackage{graphics} 
\usepackage{txfonts}
\usepackage{times}    
\usepackage[pdftex]{hyperref}
\usepackage{color}
\usepackage{textcomp}
\usepackage{booktabs}
\usepackage{ccicons}
\usepackage{todonotes}
\usepackage[ampersand]{easylist}
\usepackage{multirow}

\title{Online Group-exercises for Older Adults of Different Physical Abilities
}

\author{
\IEEEauthorblockN{
Marcos Báez\IEEEauthorrefmark{1},
Francisco Ibarra\IEEEauthorrefmark{1},
Iman Khaghani Far\IEEEauthorrefmark{2}, 
Michela Ferron\IEEEauthorrefmark{3} and
Fabio Casati\IEEEauthorrefmark{1}}

\IEEEauthorblockA{
\IEEEauthorrefmark{1}
University of Trento\\
Via Sommarive 9, Trento, Italy\\
}

\IEEEauthorblockA{
\IEEEauthorrefmark{2}
Northeastern University \\
360 Huntington Ave, Boston, MA 02115, United States \\
}

\IEEEauthorblockA{
\IEEEauthorrefmark{3}
Fondazione Bruno Kessler \\ 
Via S. Croce 77, Trento, Italy \\
}
}

\begin{document}

\maketitle

\begin{changemargin}{0cm}{0cm} 
\begin{abstract}

In this paper we describe the design and validation of a virtual fitness environment aiming at keeping older adults physically and socially active. We target particularly older adults who are socially more isolated, physically less active, and with less chances of training in a gym. 
The virtual fitness environment, namely Gymcentral, was designed to \emph{enable} and \emph{motivate} older adults to follow personalised exercises from home, with a (heterogeneous) group of remote friends and under the remote supervision of a Coach. We take the training activity as an opportunity to create social interactions, by complementing training features with social instruments.  
Finally, we report on the feasibility and effectiveness of the virtual environment, as well as its effects on the usage and social interactions, from an intervention study in Trento, Italy.\footnote{
\copyright2016 IEEE. 
To be published in The 2016 International Conference on Collaboration Technologies and Systems (CTS 2016).

Personal use of this material is permitted.
Permission from IEEE must be obtained for all other uses, in any current or future media, including reprinting/republishing this material for advertising or promotional purposes, creating new collective works, for resale or redistribution to servers or lists, or reuse of any copyrighted component of this work in other works must be obtained from the IEEE.

For more details, see the \href{http://www.ieee.org/publications_standards/publications/rights/copyrightpolicy.html}{IEEE Copyright Policy.}

}
  
\end{abstract}
\end{changemargin}

{\keywords virtual gym, older adults, social interactions, intervention study}

\section{Introduction}

Engaging in physical activity can bring multiple benefits to the health and well-being of older adults \cite{spirduso2001exercise}. It reduces risk of falls \cite{thibaud2012impact}, slows progression of degenerative diseases \cite{stuart2008community}, and even improves cognitive performance and mood \cite{landi2010moving}. Sedentary behaviour, on the other hand, is associated with negative effects such as increased risk of diabetes, cardiovascular disease, and cardiovascular and all-cause mortality \cite{wilmot2012sedentary}. 


However, a variety of barriers make engagement in regular physical activity difficult for older adults: lack of  adequate facilities and infrastructures, reduced functional abilities, lack of motivation \cite{schutzer2004barriers}, and in general, simply because it is no longer easy to leave their homes and participate in physical activities on a regular basis.
Thus, and in spite of the growing evidence of the benefits of physical activity, as well as the adverse effects of sedentary behaviour, physical inactivity is still prevalent in older adults \cite{harvey2013prevalence}. 

For similar reasons, older adults - especially those with decreased mobility - have more limited opportunities to engage in social activities. This reduced participation in social activities, along with changes in social roles, puts them at risk of social isolation, which has been associated with negative effects in physical and mental health \cite{bower1997social}. Engaging in social interactions is, ironically, one important factor in helping to counter these effects \cite{fratiglioni2000influence}.

Home-based interventions for physical training have the potential to overcome these limitations and enable the homebound to engage in physical and social activities \cite{stuck2002home}. Current solutions provide effective support for the general population, especially for outdoor activities and for people that do not require expert coaching \cite{farFitness2016}. However, as we will see in the paper, sensible groups such as older adults find less support, with solutions not coping with their specific needs, motivational drives, and social context.

In this paper we present a tablet-based fitness environment, namely Gymcentral, designed to keep independent-living older adults physically and socially active. We do this by providing trainees with a virtual environment that is both \emph{personal}, i.e., the training program and feedback are personalised, and \emph{social}, i.e., members can interact and participate to group exercise sessions even if they have different physical abilities. The application is built on years of research on home based-training
\cite{silveira2013motivating, silveira2013tablet, far2014virtual}, and has been shown to enable and motivate older adults to exercise \cite{far2015interplay}. In this paper, we focus on the design aspects, with the following contributions: 

\begin{itemize}
\item identification of relevant \emph{design dimensions} for technology seeking to provide online group exercising to older adults, including groups with different physical abilities

\item implementation and evaluation of a \emph{virtual fitness environment} that builds on the design dimensions and principles identified, in a physical home-based intervention study 
\item a qualitative study of \emph{online social interactions} resulting from a training context, which provides insights to improving social features in online fitness environments  
\end{itemize}

In what follows we describe the related literature, the design rationale and the results from the intervention study.

\section{Related Work}
\label{sec:relwork}



\subsection{Home-based interventions}
Physical intervention programs in the form of group-exercise sessions or home-based training have shown equivalent physical outcomes \cite{freene2013physiotherapist}. However, group-based interventions have shown to achieve higher levels of participation in the long-term \cite{van2002effectiveness}, while in the short-term the results are comparable or still not conclusive \cite{van2002effectiveness,freene2013physiotherapist}.  

The evidence in favour of group-based exercising can be explained by the importance of socialising as a motivating factor in physical training \cite{phillips2004motivating,de2011older}. A study by de Groot \cite{de2011older} reported that older adults do indeed prefer training with others rather than individually. 
However, group exercising might be a challenging (or infeasible) setting for older adults due to their heterogeneity. In particular, different levels of skill between participants might result in motivation problems and, consequently, affect the effectiveness of the exercises \cite{de2011older}. This and other obstacles that older adults experience, such as reduced mobility, makes of home-based individual interventions the only option for some older adults to attend to group exercises.


\subsection{Technology for home based-training}


DVDs \cite{Wojcicki2014} and tablets \cite{Silveira2013a} have been used to facilitate home-based training for older adults, and increasingly, gaming technology.
For example, the Wii has been used both as customised \cite{Carmichael2010} and off-the-shelf \cite{Agmon2011} solutions to train balance or physical activity in general. Nonetheless, none of these focus on virtual group-exercising, and moreover, only a few have been tested by older adults at home \cite{Agmon2011,Silveira2013a,Wojcicki2014}.

Virtual environments (VE) are also commonly used in these systems. Older adults prefer technologies that have familiarities with their everyday life \cite{Ijsselsteijn2007a,ThengYin-leng;ChuaPuay-Hoe;Pham2012} and show preference towards VE over video tutorials \cite{Waller1998}.  In this sense virtual environments can better represent the real-life experience of a gym. Nonetheless, most solutions are focused on the physical training experience and overlook other aspects such as the feeling of training together with others.

\subsection{Persuasion technology}

Persuasion strategies \cite{Oinas-Kukkonen2008} have been used in home-based training applications in order to motivate older adults to increase training duration and adherence. Strategies can be categorised in two ways: as individual, those that do not need a social community (i.e. reminders and suggestions, positive and negative reinforcement, self-monitoring and rewards); and social, including motivation instruments that are leveraged by social interaction.

While individual strategies have been tried \cite{Rodriguez2012}, older adults seem more inclined to applications that enable social interactions \cite{Brox2011,Ijsselsteijn2007a}. Studies reveal that older adults are more interested in exercises that provide healthy competition and collaboration \cite{Ganesan2012}, and prefer to socialise with their friends while performing similar activities \cite{Vargheese2013}. Successful implementations include a tablet-based exercise intervention \cite{Silveira2013a}, which found that older adults adhere longer to a training plan leveraging on social strategies such as family support, competition and collaboration.

\subsection{Coaching and tailoring}


Coaching can be an essential part of training: \emph{before} training, to identify trainees' needs, abilities and goals, and to prescribe tailored training plans; as well as \emph{during} and \emph{after}, to provide support, monitor progress and adjust training plans accordingly \cite{Chi-Wai2011}. Technology in the form of virtual coaching, and sensors (e.g. Fitbit, Nike\texttt{+}) can help to monitor performance and as a result provide feedback and establish new training goals.

Virtual coaching can provide the support that older adults require when exercising \cite{Chi-Wai2011}. Solutions that include coaching have achieved longer adherence times than those that do not \cite{Watson2012}, personal feedback has lead to improved accuracy of exercising and performance, by giving trainees a better understanding of the instructions \cite{Qian2010}. 
Nonetheless, however useful these tools might be, they cannot replace a human coach yet. Human coaching has been shown to provide better emotional and psychological support during training, and it is still required for risk assessment and tailoring of training \cite{Chi-Wai2011,Hanneton2009}.


\section{A VIRTUAL GYM FOR OLDER ADULTS}
Designing fitness applications to \emph{enable} and \emph{motivate} independent-living older adults - of potentially different abilities - to follow group exercises from home, pose many design challenges. In this section we follow the design process of Gymcentral - an application that addressed the aforementioned scenario - to motivate the usage scenario and provide design recommendations.




\begin{figure*}
  \includegraphics[width=\textwidth]{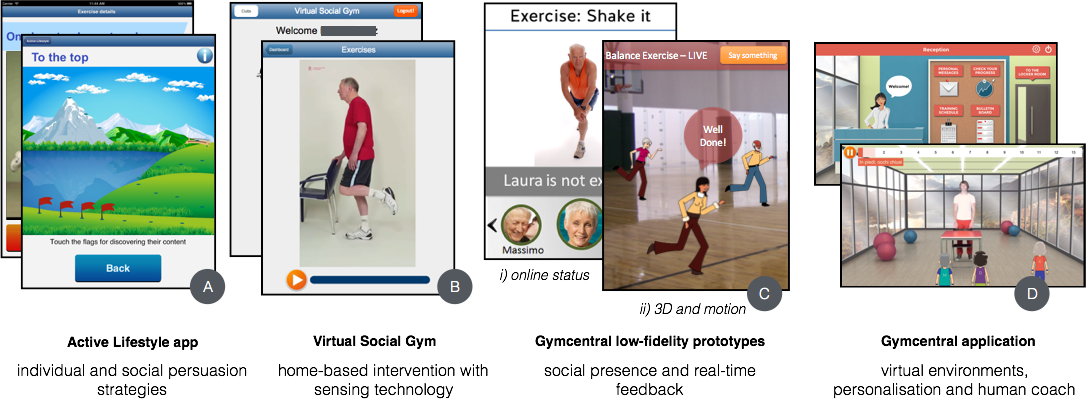}
  \caption{Training applications. a) Active Lifestyle app, exploring the use of individual and social persuasion strategies; b) Virtual Social Gym, exploring the use of activity monitors in home-based interventions; c) Gymcentral early design alternatives; d) Gymcentral application in its current form. }~\label{fig:figure1}
\end{figure*}

\subsection{Design space and rationale}
The design of Gymcentral as a tool for online group exercising is informed by evidence in the literature (see Section \ref{sec:relwork}) as well as previous experiences that progressively shaped the current implementation of the application. 
 
\emph{Active Lifestyle} (Figure \ref{fig:figure1}a) explored the feasibility of providing a home-based strength and balance exercise program by means of video exercises in a tablet device \cite{silveira2013motivating}. In addition, it studied the effects of using individual (e.g., positive and negative reinforcement) and social persuasion strategies (e.g., collaboration and competition) in the adherence to the training programs \cite{silveira2013tablet}. 
This experience suggests that i) tablet-based physical interventions for independent-living older adults are feasible, ii) persuasion strategies have a significant positive effect on adherence, and that iii) social persuasion strategies are more effective than individual strategies in motivating older adults to exercise.

The \emph{Virtual Social Gym} (Figure \ref{fig:figure1}b) application added domain knowledge from training experts to provide tailored home-based exercise programs to independent-living older adults. This application allowed the training expert to define training programs, along with training profiles corresponding to different levels of intensity, and monitor the progress of users along the training program. 
Sensors collected user activity data, which was presented to the expert in a web-based dashboard. Results from this project i) stressed the importance of tailoring exercise programs, ii) reinforced previous studies suggesting the importance of a human coach, and iii) confirmed the feasibility of performing remote monitoring by employing an activity monitor in the context of a home-based physical intervention (full study protocol in \cite{geraedts2014adherence}). 

From these previous experiences and literature we derive the following main dimensions and related recommendations:

\begin{easylist}
\ListProperties(Hide=100, Hang=true,Style*=--~)

& \emph{Tailored training and feedback}. Tailoring a training program is an essential part of the coaching process \cite{Chi-Wai2011}, and as such should be incorporated in the design. It involves assessing the abilities of the trainee and constantly tuning the program so that it remains both safe and effective \cite{geraedts2014adherence}. 

& \emph{Human expert in the loop}. Coaching either by real or virtual coaches can be more effective and motivating than no coaching for the trainees at home \cite{ijsselsteijn2004virtual}. However, when dealing with the older population, studies emphasise the need for a real coach \cite{Hanneton2009,geraedts2014adherence}.


& \emph{(Social) Persuasion Strategies}. Self-efficacy (i.e., perceived capability and confidence), a strong predictor of adherence to physical exercises, is less exhibited in older adults compared to populations of different age groups \cite{phillips2004motivating}.
Studies have shown that the use of persuasive features (especially social persuasion strategies) increases the adherence to training programs \cite{silveira2013tablet}.

& \emph{Social Interactions}. Engaging in activities with others can help stimulate social interactions \cite{leonardi2008supporting}. This is particularly beneficial for older adults with limited opportunities to interact - in most cases for the same reasons they need home training. Training together could then potentially help older adults to stay physically and socially active. 

\end{easylist}

In this paper we're addressing the additional challenge of enabling older adults of different abilities, and despite this difference, to engage in \emph{group exercises} from home. Providing this experience poses extra design requirements that were not addressed in the aforementioned works. Thus, we explored different design alternatives to realise the group exercising (see Figure \ref{fig:figure1}c.), from simply indicating that another trainee was also training (online status) to having a real-time motion and visualisation (3D and motion), each alternative with a different level of immersion, feedback and requirements in terms of technology.

The design alternative materialised in the current version of the tool (Figure \ref{fig:figure1}d), relies on the following design aspects:

\begin{easylist}
\ListProperties(Hide=100, Hang=true,Style*=--~)
& \emph{Virtual environments}. Virtual environments have been shown to increase the sense of presence, or psychological immersion \cite{grinberg2014social}. 
& \emph{Social presence and privacy}. Social presence, along with user embodiment (avatars), help to reduce physical barriers and get users more engaged in the activities while preserving their privacy \cite{siriaraya2014exploring}.
& \emph{Keeping disparities invisible to the group}. Avatars do not follow the actual trainee's movements but predefined movements. This was both a practical constraint (i.e., to keep the technological requirement to a minimum) and a design constraint (i.e., to keep the specifics of the exercise performed hidden from others) to avoid the negative effects of face-to-face group exercising \cite{de2011older}. 
\end{easylist}



Gymcentral thus relies on the metaphor of a \emph{virtual gym} for the added benefits explained before, as well as to compensate for the added complexity. The specifics of this version are discussed below.

\begin{figure*}
\centering
  \includegraphics[width=\textwidth]{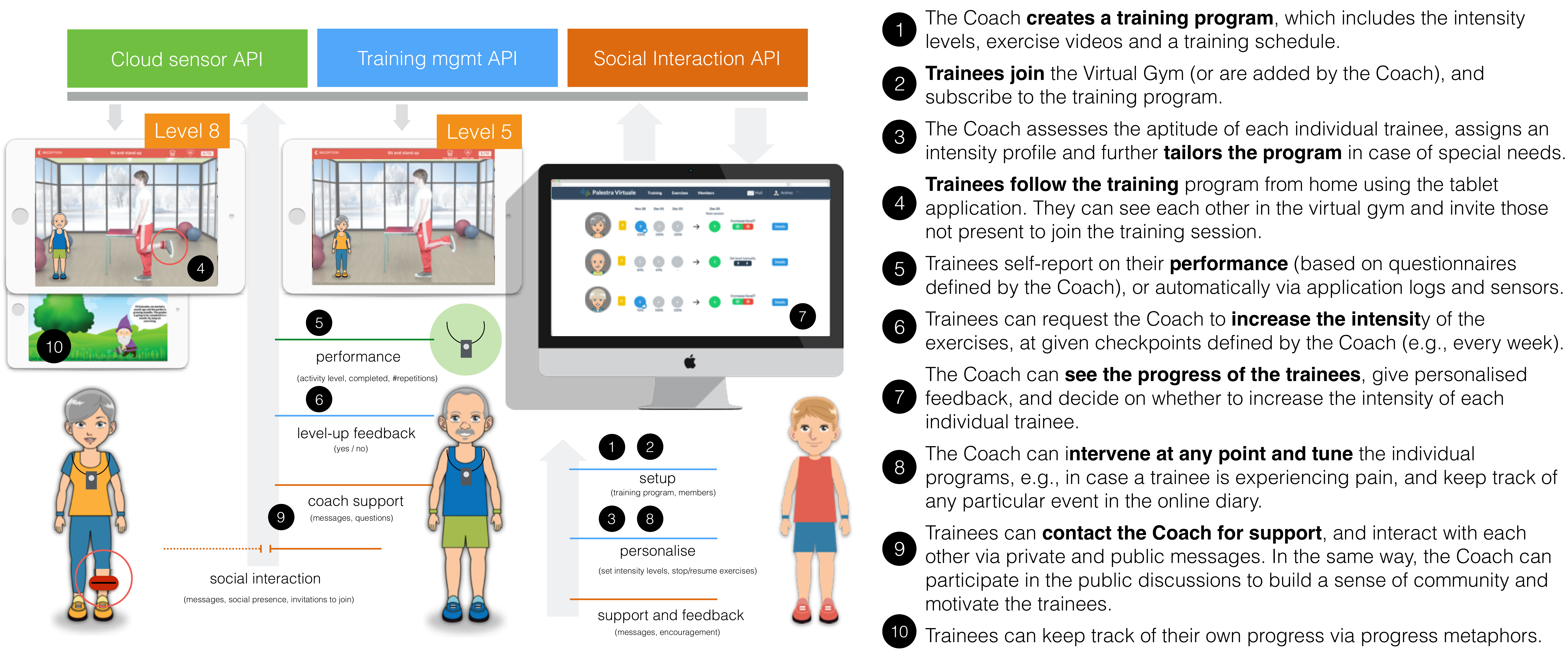}
  \caption{Overview of the Gymcentral service.}~\label{fig:figure2}
\end{figure*}

\subsection{Gymcentral Applications}
The Gymcentral platform is organised in two main applications that serve the needs of both trainees and the coach. Together, these applications can support a typical workflow as illustrated in Figure \ref{fig:figure2}.

\subsubsection{Trainee's Application}

\begin{figure*}
\centering
  \includegraphics[width=\textwidth]{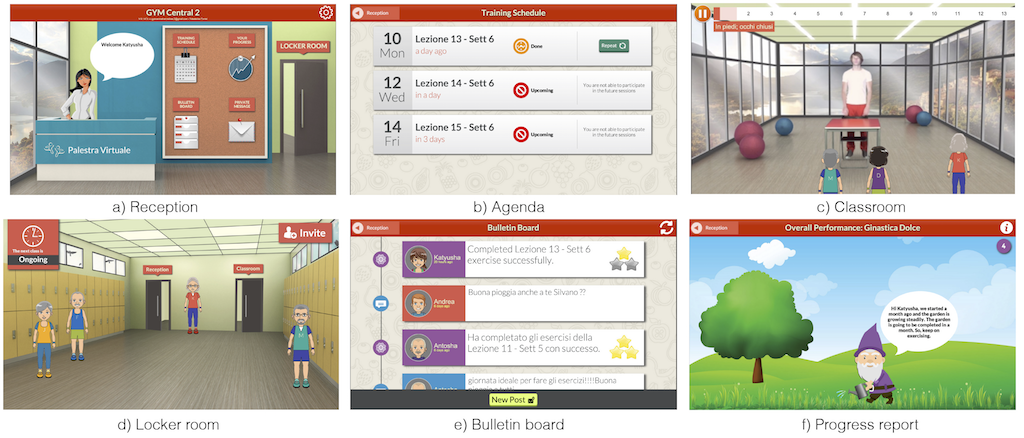}
  \caption{Main features of the Trainee Application}~\label{fig:features}
\end{figure*}

It allows the trainees to follow tailored training programs from home, unassisted, using sensors and a tablet device. The design of this application relies on the metaphor of a \emph{gym}, providing similar spaces and services (Figure \ref{fig:features}):





\begin{easylist}
\ListProperties(Hide=100, Hang=true,Style*=--~)
& \textit{Reception.} 
The entry point of the Gym, where the user has access to all the services. A virtual receptionist helps the user in getting oriented, e.g., informs of new messages and upcoming sessions.

& \textit{Locker Room.} 
A space where trainees usually meet each other and get ready for the training classes. In the locker room, users can see each other (as avatars), interact by means of predefined messages (e.g. ``Hi, let's go to the classroom"), and invite members who are not online to join. 

& \textit{Classroom.} 
 A space where users have access to the exercise instructions (video blended with the gym environment). Users in the classroom can see the coach as well as the other trainees (as avatars). 

& \textit{Agenda.} It displays the training schedule of the current week, highlighting user participation to the sessions. 

& \textit{Messages}. The bulletin board is a community feature where trainees can exchange public messages. Performance and exercise achievements of the trainees are also automatically published on the bulletin board. Private messaging allows users to interact one-to-one with other trainees and the coach. 

& \textit{Progress report.} It displays the progress of the trainee in the training program by means of a growing garden metaphor. 
\end{easylist}

In the above sections, the Trainee app implements \emph{persuasive strategies} (e.g., self-monitoring via progress reports), \emph{social interactions} (e.g.,  messages and real-time interactions), and allows for \emph{Coach feedback} (e.g., via private messages).

\subsubsection{Coach Application}
A web-based application allows training experts to run a \emph{virtual gym}, providing support for the training and community aspects with the following features:

\begin{easylist}
\ListProperties(Hide=100, Hang=true,Style*=--~)
& \textit{Community building}. 
To start a virtual gym from scratch, defining and managing the members of the community (trainees and other coaches). Public (bulletin board) and private channels (messages) are in place to help the coach engage the trainees and build a sense of community.



& \textit{Definition of training activities}.
To organise the training activities (video exercises augmented with metadata) around fitness classes, targeting groups of users with similar needs but different abilities. The coach can associate different intensity levels and performance indicators (e.g. measured with questionnaires or sensors) to these training classes.

& \textit{Initial assessment}. 
The coach can define pre-assessment exercises (and require specific aptitude information) before accepting the trainee, and then use this information to set a starting intensity level. Special requirements can be logged in the online diary and used to tailor the exercises. 

& \textit{Monitoring}. 
To continuously monitor the progress of trainees. The monitoring covers as a minimum the participation of trainees and the completeness of the exercises, and can incorporate the performance indicators defined by the coach (self-reported data and measures from sensors).

& \textit{Personalisation and safety in the training program.} 
 To increase the intensity level of individual trainees based on their performance (according to the pre-defined intensity profiles). In addition, the coach can tune the program to stop (and eventually resume) individual exercises, for example, in case of injuries.


& \textit{Personalised feedback.} Reports facilitate the process of looking at the performance of trainees, and providing personalised feedback in context. Feedback is sent to trainees using the private message feature.
\end{easylist}

\section{METHODS}
In this section, we explain the objectives of the study, its participants and the intervention design.

\subsection{Objectives}
The study reported in the paper was performed in the context of a larger intervention study aiming at evaluating the feasibility of the technology and its effects on physical wellbeing. In this study we focus on the former, investigating the following design aspects related to the virtual group exercising:

\begin{easylist}
\ListProperties(Hide=100, Hang=true,Style*=--~)

& \emph{usability and learnability of the application by older adults}; 
aiming to understand the usability at the beginning and at the end of the intervention, especially in relation to its complexity.

& \emph{acceptance of the technology by older adults},  
looking at different subjective aspects of the experience with technology as a whole.

& \emph{perceived value of main design dimensions}, 
aiming to understand the usefulness of the proposed features as perceived by the users.

& \emph{nature of social interactions} originated within the system, looking especially at the emerging themes in the conversations in different scenarios.

\end{easylist}

In the following subsections we describe the study design and measures used to elaborate these aspects.


\subsection{Participants}

Participants aged 65+, self-sufficient and with a non-frail, transitionally frail, or a mild frailty level were considered eligible for the study. Frailty level was measured using the Groningen Frailty Indicator \cite{steverink2001measuring}, a validated questionnaire that screens for self-reported limitations in older adults. 

A total of 40 participants between 65 and 87 years old were recruited through two local volunteering organizations (29 females and 11 males, mean age = 71, s.d. = 5.7). All participants obtained a formal written approval by their family doctor to allow them to participate in the study. Both doctors and participants received a written outline and explanation of the study before participating. 

Five participants withdrew at different times during the course of the study due to unpredictable health or family problems. One participant was substituted because the withdrawal occurred before the beginning of the study, while the others could not be replaced since they withdrew during the course of the study.
Results are therefore based on the data from 36 participants (27 females and 9 males, mean age = 71.2, s.d. = 5.8, between 65 and 87 years old).


\subsection{Study design}



The study followed a framework for the design and evaluation of complex interventions in health settings \cite{campbell2000framework} and lasted for a total of 10 weeks, including one week at the beginning for technical deployment, application testing and the collection of initial questionnaires, and one week at the end for the administration of the final questionnaires. Using a matched random assignment procedure based on age and frailty level, participants were assigned to either an experimental (social) or control condition.

Participants of the social group were assigned the full version of the Trainee App, including the personalised program along with the social and persuasion features (the condition with the more complex set of features). The control group, instead, had a basic version of the application, which included the personalised exercise program, but no persuasive, social and self-monitoring features.

In the social condition, participants could communicate with each other and with the coach using the messaging features of the application (bulletin board and private messages), while participants in the control group could do so by telephone. In order to mirror the time and attention provided to the social group (able to communicate with the coach through the application), periodic telephonic contact was maintained with the control group by a community manager \cite{michaelinterventions}. 

Prior to the beginning of the intervention, participants took part in a workshop to learn how to use the tablet and the Gymcentral application, and were provided with handouts containing information about the study, the use of the tablet and of the application. Additionally, each participant received a necklace sensor that included a 3D accelerometer and a barometric pressure sensor in order to monitor their physical activity. They also participated in individual sessions with the personal trainer, who assessed their physical health and ability, and assigned them an initial training level. 

The intervention consisted in 8 weeks of physical training based on the Otago Exercise Program, specifically tailored for older adults \cite{gardner2001practical}. The training program consisted of 10 levels of increasing intensity, which included simple exercises based on functional everyday movements. During the exercise program, participants were asked to perform at least two training sessions per week. In both social and control groups, level-up was gradually suggested every week by the application. If participants agreed to level-up, the following level was unlocked, requiring a confirmation from the personal trainer through the Gymcentral coach application in the case of the social group. 


The study received ethical approval from the CREATE-NET Ethics Committee on ICT Research Involving Human Beings (Application N. 2014-001). 

\subsection{Measures}
\subsubsection{Usability}
The usability 
was assessed using the System Usability Scale \cite{brooke1996sus}, a 10 item questionnaire with five response options (1 = completely disagree, 5 = completely agree), at two time points: at the beginning of the study (after the tutorial on the application, when participants used the application for the first time), and at the end of the study. 
These measures were obtained for both groups in order to compare the usability of the different interface complexity levels. 


\subsubsection{Technology acceptance}
To evaluate technology acceptance, we developed a questionnaire on the basis of previous literature \cite{phang2006senior} investigating the following dimensions: \emph{anxiety} towards Gymcentral, \emph{attractiveness} and acceptance of the application, \emph{satisfaction} of the service provided and perceived \emph{usefulness} of the application. Participants expressed their preferences on a 5-points Likert scale (1 = completely disagree, 5 = completely agree). We expected these dimensions to improve after the training program.

\subsubsection{Usefulness by feature}
To analyse usefulness by feature we provided a short questionnaire\footnote{\url{https://goo.gl/zl7daL}}, asking participants to report on a 5 point Likert scale how useful they though each feature to be. 


\subsubsection{Nature of social interactions}
To investigate the nature of social interactions within the application, we performed a qualitative analysis of both the messages posted to the bulletin board and private messages. We developed a coding scheme in two steps: first, we categorized the messages without using pre-existing categories, then we compared our classification to those provided in the relevant literature about online behaviour and communities (e.g., \cite{pfeil2007patterns}), developing a final coding scheme composed of 5 top- and 12 sub-categories. 

\section{Results}


\subsection{Usability and technology acceptance}
%

Pre- and post- scores of usability and technology acceptance for the social and control group are shown in Table~\ref{table:usaacc}.

\subsubsection{Usability}
A mixed between-within subjects analysis of variance was conducted to compare pre- and post- scores of the System Usability Scale between participants in the experimental and in the control group. The analysis showed a significant interaction between group and time (F(1, 34)= 8.286, p = .007), and a significant main effect for time (F(1, 34)= 37.113, p $<$ .001), and for group (F(1, 34)= 14.614, p = .001). This suggests that, while at the beginning there was a noticeable difference in the usability of the two Gymcentral versions, with the basic version performing better, the usability of the full application improved significantly more over time in the social group than in the control group. Although initially the full application was reported as more difficult to use, at the end of the study its perceived usability increased to reach a level comparable to the one of the basic version.


\begin{table}
  \caption{Usability and technology acceptance pre- and post- measures (range: 1 to 5)}
  \small
  \begin{tabular}{| l | l | l | l | l | }
    \hline
    \multirow{2}{*}{} &
      \multicolumn{2}{c}{Control} &
      \multicolumn{2}{c|}{Social} \\
    & Pre (Err) & Post (Err) & Pre (Err) & Post (Err) \\
    \hline
    Usability & 4.25 (.16) & 4.62 (.11) & 3.33 (.13) & 4.36 (.13) \\ \hline
    Anxiety & 1.44 (.24) & 1.66 (.25) & 2.40 (.25) & 1.78 (.16) \\ \hline
    Attractiveness & 4.06 (.24) & 4.53 (.18) & 3.85 (.24) & 4.50 (.16) \\ \hline
    Satisfaction & 4.44 (.20) & 4.60 (.14) & 4.13 (.23) & 4.65 (.09) \\ \hline
    Usefulness & 4.50 (.15) & 4.56 (.18) & 3.93 (.24) & 4.65 (.13) \\ \hline
  \end{tabular}

  \label{table:usaacc}
\end{table}

\subsubsection{Technology acceptance}
A mixed between-within subjects analysis of variance was conducted to compare each of the following dimensions of technology acceptance at the beginning and at the end of the study.

\textbf{Anxiety.} The analysis showed no significant interaction between group and time (p = .069) and no significant main effect for time (p = .372), but showed a significant main effect for group (F(1, 34)= 5.543, p = .024). A closer analysis of the studentized residuals 
allowed us to detect two outliers in the data. After removing those observations, the analysis showed a significant interaction between group and time (F(1,32) = 4.713, p = .037), suggesting that anxiety towards the application \emph{significantly decreased over time for the social but not for the control group}. 

\textbf{Attractiveness.} The analysis did not reveal a significant interaction between time and group (p = .661), nor a significant main effect for group (p = .576), but it showed a significant main effect for time (F(1,34) = 7.448, p = .01). Multiple comparison tests with Bonferroni correction showed that attractiveness of the application significantly increased for the social group (p = .023) but not for the control group (p = .134). While the results of the analysis of variance suggests that, taken together, both groups reported to like the 
trainee application significantly more at the end of the study, post-hoc comparisons suggest that this \emph{difference was significant for the social group but not for the control group}. 

\textbf{Satisfaction.} The analysis did not reveal a significant interaction between time and group (p = .308), nor a significant main effect for group (p = .49) or for time (p = .051). However, multiple comparison tests with Bonferroni correction showed that \emph{satisfaction significantly increased from pre- to post- for the social} (p = .028) \emph{but not for the control group} (p = .513). Furthermore, an exploration of the studentized residuals revealed the presence of one outlier. The analysis of variance repeated after excluding the outlier showed a significant main effect for time (F(1,33) = 6.561, p = .015). 

\textbf{Usefulness.} The analysis did not reveal a significant interaction between time and group (p = .09), nor a significant effect of the main effect for group (p = .192), but it showed a significant main effect for time (F(1,34) = 4.291, p = .046). Multiple comparison tests with Bonferroni correction suggest that there was a significant increase in the perceived usefulness of the application for the social (p = .007) but not for the control group (p = .827). Consistently with the previous analyses, this suggests that, overall, participants perceived Gymcentral as more useful after the training program, and that \emph{perceived usefulness in the social group may have improved more with respect to the control group}.

\subsubsection{Discussion}
Not surprisingly, the usability of the application was lower for the Social group at the beginning of the study, reflecting participants’ initial difficulties to deal with a more complex user interface. However, by the end of the training program usability had increased significantly, approaching the top end of the scale. We should further investigate the role played by the use of the metaphor of the virtual environment in the learnability of the application.
Overall, while Internet connection was an intermittent issue, both usability and technology acceptance of the Trainee application (both versions) generally improved. 
For the full application, these results mean that 
users could handle the extra complexity and learn to use this type of tool.

\subsection{Trainee's feedback on the features }

\subsubsection{Usage and perceived usefulness}
\begin{figure}
\centering
  \includegraphics[width=\columnwidth]{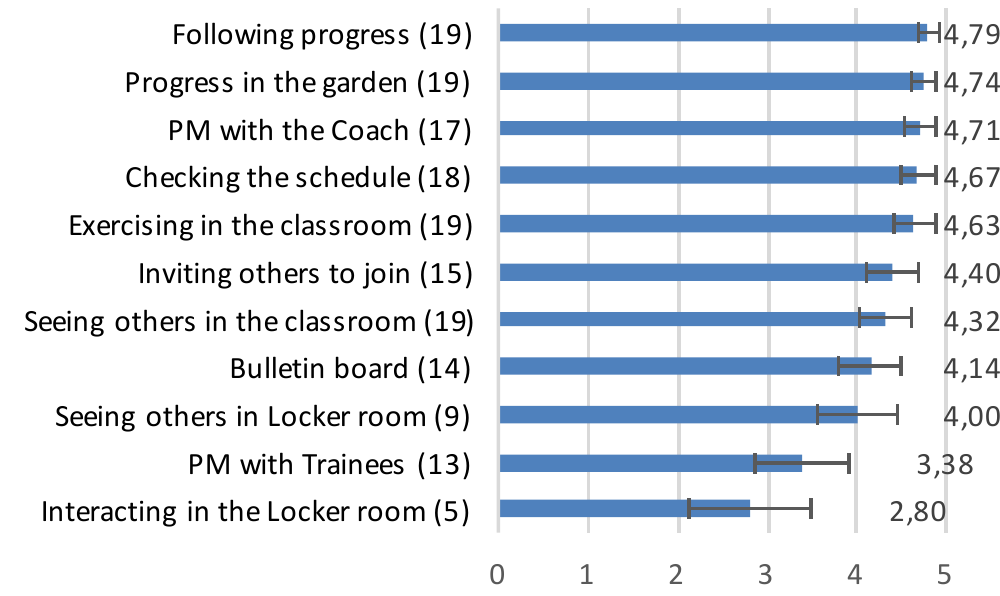}
  \caption{Perceived usefulness of each of the features. The number of users that experienced the feature is indicated in parentheses.}~\label{fig:figureAll}
\end{figure}

In order to understand the value of design dimensions and recommendations that we have identified, we asked participants of the Social group - which were assigned the full-featured version of the app - to report on the usage and perceived usefulness of the features of Gymcentral. The results are illustrated in Figure \ref{fig:figureAll}. 

The features that are instrumental to the training were naturally experienced by most of the trainees, and this includes 
\emph{exercising in the classroom} and \emph{checking out the schedule}. What is interesting is that \emph{training in company} was also experienced by most trainees. Together, these features enabling the group training were highly regarded by trainees. 

Persuasion features were also among the most experienced and valued. This includes, 
\emph{following the progress} and visualising their own \emph{progress in the garden} and, still very positive but to a lesser extent, \emph{inviting others to join} a training session. 

Social interaction features received mixed results. The most useful and experienced feature was \emph{private messaging with the Coach}, followed by the public messages in the \emph{bulletin board}. Interestingly, \emph{private messages with other trainees} were perceived as less useful, indicating a higher preference of trainees for interaction with the entire group rather than individually. We expand on the nature of these interactions in the next subsection.  

While the \emph{social presence in the Classroom} was highly rated, participants regarded the features
present in the \emph{Locker room} among the least useful. The Locker room was designed as a place for trainees to meet and socialise before starting a training session. They would invite others, wait for them before starting the session, and in the meantime interact via predefined real-time messages. In practice, the user behaviour was different. 
The application logs show that users were not waiting for others in the Locker room after sending their invitations, instead they would go directly to the Classroom and wait there for others to join.
\subsubsection{Positive and negative aspects}

Participants from both groups were asked to provide feedback about the positive and negative aspects of the experience. 

\textbf{What aspect was the most fun and motivating?} In the Control group, the topics that dominated the feedback were the possibility of training from home (``Being able to exercise at any time, and from my living room"), personal satisfaction (``Satisfaction of performing the exercises every day") and discipline (``The personal commitment to perform the exercises"). Interestingly, one participant reported the physical meetings as the most fun part (``The meetings with the project organisers").

In the Social group the dominant aspect was the social features, with participants citing the possibility of exercising with others (``Feeling that you're training with others and followed by the Coach"), being invited to join (``Being invited to exercise in the virtual gym"), 
and messaging with other participants and the Coach (``Very nice to find messages in the bulletin board"). As in the previous group, one participant reported also the initial meetings as one of the highlights.

\textbf{What aspect did you like the least?} Both groups reported the same negative aspects regarding the experience. The dominant aspect was the problems with the application, which were due to Internet connectivity issues in some areas of the city (``When [the training] was not loading"). 
We highlight the feedback from one participant of the Social group,
who reported to have gone to a friend's house to use Internet (``The tablet was not working at home, so I had to go to a friend's house to exercise").
Another issue reported was the monotony of exercises (``I've found the exercises repetitive") which is probably related to the long term nature of the training. 

\subsubsection{Discussion}
The user feedback \emph{reinforces the value of group exercising, social interactions and persuasion} features of the application, and of the design recommendations on which it is built. The results also point to i) the need for more effective mechanisms to motivate social interactions among community members, as shown by the perception and usage of messaging features, and ii) the need for more effective environments for motivating real-time social interactions, as shown by the user feedback on the Locker room feature.

\subsection{Analysis of online social interactions}


\begin{figure*}[tb]
\centering
  \includegraphics[width=\textwidth]{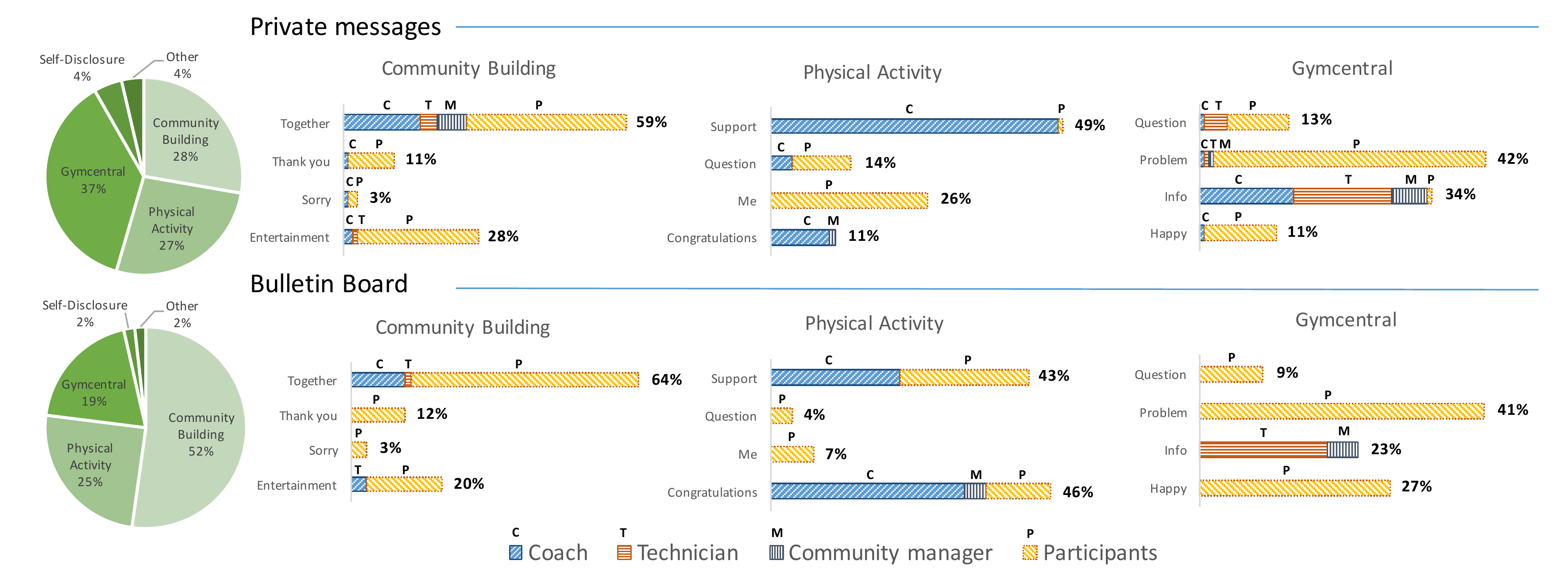}
  \caption{Nature of social interaction in private and public messages.}~\label{fig:figure5}
\end{figure*}

Private messages were preferred (411 messages) over the bulletin board (133). To better understand the type of messages exchanged, a manual classification was done. A 20\% random sample of all messages was coded manually, initially without using pre-existing categories, and later coded with the scheme detailed in Table \ref{table:coding}. The coding scheme was developed based on relevant behaviour and communities literature (e.g., \cite{pfeil2007patterns}).

\begin{table}[!ht]
\caption{Messages coding scheme}
\small
\begin{center}
    \begin{tabular}{ |p{0.99\columnwidth}| } 
    \hline    
    \textbf{Community building}  \\ 
    \hline 
    \textit{Togetherness.} Interacting with the community or particular members, including invitations and messages of welcome, to stimulate participation, and to stress the value of the community. \\

\textit{Thank you.} Thanking the community or a member for the help, support or for their understanding. \\

    \textit{Sorry.} Apologising for an action.\\

    \textit{Entertainment.} Sharing jokes, quotes or aphorisms. \\
    \hline    
    \textbf{Gymcentral Application}  \\ 
    \hline                    
\textit{Satisfaction.} Sharing a positive experience with the application or the study.

\textit{Problem.} Reporting problems with the application (e.g. internet connection or technical issues).

\textit{Information.} Providing information or announcements about the application. Giving advice, recommendations, and suggestions.

\textit{Question.} Asking for information about the app and technical issues. \\
    \hline    
    \textbf{Physical Activity}  \\ 
    \hline

\textit{Support.} Offering advice, support or sympathy to the community or a particular member. Encouraging others to participate.

\textit{Congratulations.} Congratulating the community or a particular member for participating in the program or completing an exercise.

\textit{Me.} Sharing personal experience on the training (e.g. level of commitment, participation, level-up intention, problems).

\textit{Question.} Asking for information about the training or exercise performance. Requesting for a level-up.\\

    \hline    
    \textbf{Self-disclosure}  \\ 
    \hline   
Sharing personal experience or information not related to the training (e.g. personal stories, daily activities)  \\  
    \hline    
    \textbf{Other}  \\ 
    \hline       
     Messages that did not fit in any of the other categories\\
    \hline     
    
    \end{tabular} 

\label{table:coding}    
\end{center}
\end{table}

Two independent coders classified the messages sampled. Cohen’s kappa coefficients for the bulletin board were .85 for top-categories and .84 for sub-categories; and for the private messages coefficients were .87 for top-categories and .85 for sub-categories, indicating a general high agreement. After the independent coding, a single coder classified all  messages and combined the results. These are shown in Figure \ref{fig:figure5}.

\textbf{Bulletin board.} 
Used mainly to promote \emph{community building}, in particular togetherness. Participants had an active role, they posted greeting messages (e.g. ``Good morning everybody!") and used a humorous tone in the conversation (e.g. ``You are a little crazy"). The bulletin board was also used to publicly thank the Coach and other participants for their help or invitations to train together. To a lesser extent, the Coach contributed to community building by welcoming and greeting participants (e.g. ``Have a nice start of the week everybody!"). 

The talk about \emph{physical activity} was centred in congratulating and offering support for the training. In particular, the Coach was very active, encouraging participants to attend to the training sessions, and congratulating them for their performance and the level-up requests (e.g. ``Well done everybody... many of you wrote me... to level-up").

The messages regarding the \emph{Gymcentral} application were mostly about technical issues. The technician used the bulletin board to broadcast advice and information on these issues. At certain points during the study, participants experienced slow connection problems that compromised the proper functioning of the application, especially the streaming of exercise videos. However, there were also positive comments about the application and the garden metaphor (e.g. ``Oh! A bright butterfly appeared in the garden, wonderful, thank you!").

\textbf{Private messages.} Most 
messages were about the \emph{Gymcentral} application. As in the bulletin board, almost half of the messages about the application were directed to the technician and the Coach to report technical issues or the inability to train because of connection problems. Participants also exchanged some positive notes about the application.

Considering the messages of \emph{community building}, we can observe that, similarly to the bulletin board, participants promoted a sense of togetherness, but the messages were more personal than the ones in the bulletin board (e.g. ``How are you?", " ...we missed you"). The more intimate nature of this channel was also used for self-disclosure. Participants talked about their lives outside of the virtual gym, even engaging in conversation with the Coach and the community manager.

In contrast to the bulletin board, when discussing \emph{physical activity}, participants did not use the private messages to congratulate and support each other. Instead, participants talked about their personal experience with the exercise, and in particular 
talked to the coach and asked for advice.

 \textbf{Role of the coach.} The coach was the most active user in both communication channels. In the bulletin board, 24 messages were posted, and regarding private messages, 120 were sent and 117 received. Interestingly, the use of the bulletin board and the private messages was different. The coach used the bulletin board to congratulate participants publicly, but private messages were sent to encourage participants to train and follow the training program.


\textbf{Discussion}. These results highlight the need for having both types of channels, since they serve very different purposes. On one hand allowing for community building in public channels, and for more personal conversations in private. We should also note that compared to other technology-based interventions, where social features (e.g., forums or social networks) were rarely used \cite{aalbers2011characteristics}, in this study the social features have been largely used by the participants. Further research is required to investigate whether this result is significant and related to the design of the tool.

\subsection{Limitations}
\textbf{Different channels for Coach support}. The interactions of the Coach with the participants were scheduled to give the same type of support. However, in  absence of social features in the version of the app used by the Control group, the support was given through phone calls. This difference in the communication channel might have introduced a potential bias in the motivation to participate. 

\textbf{Sample size and gender imbalance}. Random variability, probably due to the small sample size, might have influenced the initial difference between groups in some of the measures. We also acknowledge the gender imbalance as a potential limitation to the generalisation of the results.

\section{CONCLUSION}

In this paper we have described the design of a virtual fitness environment to facilitate and motivate older adults of different abilities to follow virtual group exercises from home. 
The application, Gymcentral, was evaluated in an intervention study - where older adults were effectively engaged in the training 
\cite{far2015interplay}


We have derived a set of useful design dimensions and recommendations for home-based training and presented an approach to group exercising that could accommodate to a heterogeneous group of older adults 
- a very common setting in this population.
The feedback from users provided insights into the usefulness of the features as well as areas for future work. 
In particular, we have seen a higher number of social interaction than previous home-based training interventions, as well as a high learnability despite the complexity of the application. We attribute these positive results to the use of the research-derived design recommendations but further research is needed for more conclusive results.
Real-time interactions, however, were not as successful. Reasons for this may be the lack of activities that would motivate users to stay in the virtual room, contextual messages not being expressive enough, or users not being used to real-time messages. 

We should note that, while related to fitness applications, the above results can also inform researchers and practitioners of social applications, including collaborative applications, targeting older adults on the aspects to consider when designing social interaction mechanisms and deciding on interface metaphors.
Indeed, as an ongoing work, we are expanding the virtual environment to include \emph{leisure spaces} and \emph{productive spaces} (volunteering), as a way to explore social interactions during purposeful activities and crowd-sourcing in virtual environments, and more importantly, to cater on the opportunities of providing tools for enabling online contributions by older adults \cite{ibarratools}.


\section*{Acknowledgment}
This project has received funding from the European Union's Horizon
2020 research and innovation programme under the Marie
Skłodowska-Curie grant agreement No 690962. This work was also
supported by the project ``Evaluation and enhancement of social,
economic and emotional wellbeing of older adults" under the agreement
no. 14.Z50.310029, Tomsk Polytechnic University.

\bibliographystyle{IEEEtran}
\bibliography{references}

\end{document}